\documentstyle[11pt,aaspp4]{article}

\begin{document}

\title{The Variable Stars and Blue Horizontal Branch of the \\
  Metal-Rich Globular Cluster NGC 6441}

\author{Andrew C. Layden\altaffilmark{1,2}}
\affil{Physics \& Astronomy Dept.,
Bowling Green State Univ., Bowling Green, OH 43403, U.S.A.}
\authoremail{layden@kracken.bgsu.edu}

\author{Laura A. Ritter}
\affil{Department of Astronomy, University of Michigan, Ann Arbor, MI
48109-1090, U.S.A.}
\authoremail{lritter@umich.edu}

\author{Douglas L. Welch\altaffilmark{1}, Tracy M. A. Webb\altaffilmark{1}}
\affil{Department of Physics \& Astronomy, McMaster University,
Hamilton, Ontario L8S 4M1, Canada}
\authoremail{welch@physics.mcmaster.ca, webb@lepus.astro.utoronto.ca}

\altaffiltext{1}{Visiting Astronomer, Cerro Tololo Inter-American Observatory. 
CTIO is operated by AURA, Inc.\ under contract to the National Science
Foundation.}
\altaffiltext{2}{Hubble Fellow.  Previous address: Department of
Astronomy, University of Michigan}

\begin{abstract}

We present time-series $VI$ photometry of the metal-rich ($[Fe/H] =
-0.53$) globular cluster NGC 6441.  Our color-magnitude diagram shows
that the extended blue horizontal branch seen in {\em Hubble Space
Telescope} data exists in the outermost reaches of the cluster.  About
17\% of the horizontal branch stars lie blueward and brightward of the
red clump.  The red clump itself slopes nearly parallel to the
reddening vector.  A component of this slope is due to differential
reddening, but part is intrinsic.  The blue horizontal branch stars
are more centrally concentrated than the red clump stars, suggesting
mass segregation and a possible binary origin for the blue horizontal
branch stars.  We have discovered $\sim$50 new variable stars near NGC
6441, among them eight or more RR Lyrae stars which are
highly-probable cluster members.  Comprehensive period searches over
the range 0.2--1.0 days yielded unusually long periods (0.5--0.9 days)
for the fundamental pulsators compared with field RR Lyrae of the same
metallicity.  Three similar long-period RR Lyrae are known in other
metal-rich globulars.  With over ten examples in hand, it seems that a
distinct sub-class of long-period, metal-rich RR Lyrae stars is
emerging.  It appears that these stars have the same intrinsic colors
as normal RR Lyrae.  Using the minimum-light color of the RR Lyrae, we
determine the mean cluster reddening to be $E(B-V) = 0.45 \pm 0.03$
mag, with a significant variation in reddening across the face of the
cluster.  The observed properties of the horizontal branch stars are
in reasonable agreement with recent models which invoke deep mixing to
enhance the atmospheric helium abundance, while they conflict with
models which assume high initial helium abundance.  The light curves
of the c-type RR Lyrae seem to have unusually long rise times and
sharp minima.  Reproducing these light curves in stellar pulsation
models may provide another means of constraining the physical
variables responsible for the anomalous blue horizontal branch
extension and sloped red clump observed in NGC 6441.

\end{abstract}


\section{Introduction}

The canonical metal-rich globular cluster ($[Fe/H] > -0.8$) has a
horizontal branch (helium core burning phase of stellar evolution)
which is compressed against the red giant branch to form a ``red
clump'' (e.g., 47 Tuc, \markcite{csw93}Carney et al. 1993; M 71,
\markcite{hod92}Hodder et al. 1992).  Standard stellar evolution
theory and synthetic horizontal branches confirm this morphology
(e.g., \markcite{ldz90}Lee et al. 1990, \markcite{dor92}Dorman 1992).
However, it is becoming clear that the situation is more complicated
than this canonical picture would have us believe.
 
First, it is evident that a significant fraction of the RR Lyrae
variable stars (horizontal branch stars lying in the pulsation
instability strip, hereafter RRL) in the field near the Sun are
metal-rich.  \markcite{gwp59}Preston (1959) first recognized this
fact, and the recent metallicity distribution of
\markcite{acl94}Layden (1994) shows that 22\% of the field RR Lyrae
within 2 kpc of the Sun have [Fe/H] $> -1.0$, and 9\% have [Fe/H] $>
-0.5$ dex.  Furthermore, the kinematics of these stars identify them
with the Galaxy's thick disk population, and a few RRL with abundances
approaching solar may be produced by the thin disk
(\markcite{al95a}Layden 1995a).
The standard horizontal branch star picture does not explain the
presence of stars hot enough to lie in the instability strip at these
ages and metallicities.

In addition, several metal-rich globular clusters contain one or more
RRL candidates (cf. \markcite{skk91}Suntzeff et al. 1991).  Note
however that the often-quoted example, V9 in 47 Tucanae, has an
anomalously high luminosity and long period compared with field RRL of
similar abundance (\markcite{csw93}Carney et al. 1993).  Nevertheless,
\markcite{al95b}Layden (1995b) showed that the number of RRL per unit
progenitor luminosity was roughly equal between the metal rich thick
disk field and globular cluster populations.  This comparison was
hampered by the incomplete search for RRL among the population of
metal-rich globulars.
 
A second important complication to the canonical picture of metal-rich
globular cluster horizontal branches (HBs) is the recent observation
of blue HBs in the cores of two globulars, and in several open
clusters.  {\em Hubble Space Telescope} observations by
\markcite{rmr97}Rich et al. (1997) showed the globulars NGC 6388 and
6441 ($[Fe/H] = -0.60$ and $-0.53$, \markcite{anz88}Armandroff \& Zinn
1988) to have long blueward extensions to their otherwise red HBs.
They initially suggested that age or dynamical effects could be
responsible for the blue HBs.  However, no deep photometry has been
published for these clusters to constrain their ages, and calculations
by \markcite{rmr97}Rich et al. (1997) indicated that stellar
interactions are too infrequent to be the culprit.  They left open the
question of the physical cause of the phenomenon.  Blue HB stars in
younger, more metal-rich open clusters have also been observed (e.g.,
NGC 6791, \markcite{knu92}Kaluzny \& Udalski 1992).
 
Theoretical studies are also focusing on the blue HB star problem.
\markcite{snc98}Sweigart \& Catelan (1998) presented three scenarios
involving enhanced helium abundance and/or stellar rotation designed
to explain the blue HB observations of \markcite{rmr97}Rich et
al. (1997).  They emphasized that observational tests including the
existence and periods of RR Lyrae variables, and the relative numbers
of HB and red giant branch stars (the $R$-ratio), could discriminate
between their scenarios.  They suggest that the blue HB extensions
seen by \markcite{rmr97}Rich et al. (1997) are extreme cases of the
``Second Parameter Effect'', in which the color distribution of stars
along globular cluster HBs is seen to depend on metallicity and a
second parameter, possibly age, Helium or CNO abundances, or core
rotation.

Thus, ample motivation exists to study the horizontal branches and RR
Lyrae content of metal-rich globular clusters.  With this in mind, we
have undertaken a variable star survey of twelve of the metal-rich
globulars listed by \markcite{skk91}Suntzeff et al. (1991) as having
poor or no variable star searches.  In this paper, we report our
results for the first cluster, NGC 6441.  If dynamical effects are
involved, the high luminosity and central concentration of this
cluster (\markcite{wmh96}Harris 1996) make it among the most likely to
harbor RRL.  This prospect is enhanced by the presence of the blue HB
stars observed by \markcite{rmr97}Rich et al. (1997).  The best
existing ground-based photometry of NGC 6441 is the photographic work
of \markcite{hnh76}Hesser \& Hartwick (1976), which reached only the
brightest stars on the horizontal branch.

In \S 2 of this paper, we describe the observation and data reduction
procedures we employed.  In \S 3 we present color-magnitude diagrams
of the cluster and nearby field.  In \S 4 we identify $\sim$50
variable stars in the cluster, derive their photometric properties,
and for some present their mean light curves.  In \S 5 we derive the
foreground reddening of NGC 6441 and discuss in detail the properties
of the cluster's RR Lyrae variables and compare them to the
predictions of the \markcite{snc98}Sweigart \& Catelan (1998) models.
We present our conclusions in \S 6, and in the Appendix we note
details on some of the variable stars.

\section{Observations and Reductions}

We obtained time-series images of NGC 6441 using the direct CCD camera
on the 0.9-m telescope at Cerro Tololo Inter-American Observatory
during two runs in May and June of 1996 (3 and 8 usable nights,
respectively). The Tek\#3 2048 CCD provided a 13.5 arcmin field of
view with 0.4 arcsec pixels.  We used filters matched to the CCD to
reproduce the Johnson $V$ and Kron-Cousins $I$ bandpasses. We
processed the raw images by following the usual procedure for overscan
subtraction and bias correction, and we used twilight sky frames to
flat-field the data.
 
In each pointing toward the cluster, we obtained a short exposure
(30--100 sec) $VI$ frame pair and a long exposure (250--300 sec) $VI$
pair.  This provided two independent magnitude estimates of the HB
stars at each observational epoch, and extended the dynamical range of
the observations.  Such pointings were obtained 0--3 times each night.
The time interval between pointings was at least 2 hours, so a typical
RR Lyrae (0.3--0.8 day periods) would undergo a significant magnitude
change.  In total, we obtained 15 pointings toward NGC 6441.  The
seeing varied between 1.2 and 3.0 arcsec (1.7 arcsec median).  Figure
1 shows a short-exposure $V$-band image of the cluster.  The bright
star to the West of the cluster was masked out in all the frames.



On a photometric night in May 1996, we obtained a $VI$ image pair of
NGC 6441, together with a large number of \markcite{aul92}Landolt
(1992) photometric standards.  We also secured an off-cluster $VI$
pair centered 13.3 arcmin North of the cluster center, outside the
tidal radius (7.8 arcmin, \markcite{wmh96}Harris 1996)
designed to determine the photometric properties of the
fore/background starfield.  The standards covered a large range in
color ($-0.3 < V-I < 4.0$) and airmass ($1.08 < X < 1.74$), and were
obtained at roughly hourly intervals throughout the night.  This time
($T$) coverage confirmed that the entire night was photometric, and
enabled us to seek and correct for small, slow variations in
extinction.  In total, 63 individual standards were observed at 13
independent epochs.  The instrumental aperture magnitudes of these
stars ($A_b$), in bandpasses $b=V$ or $I$, were fit with the equation
\[ A_b - S_b = c_{1,b} + c_{2,b} X + c_{3,b} (V-I) + c_{4,b} T + c_{5,b} T^2 
\]
where $S_b$ is the standard magnitude and $c_{i,b}$ ($i=1,5$) are the
fitted coefficients.  The time-dependent terms produced corrections of
0.01 mag or less.  The rms scatter about the adopted fits were 0.010
and 0.013 mag in $V$ and $I$, respectively.

The on- and off-cluster $VI$ pairs obtained on this night were reduced
using the DAOPHOT II point spread function (PSF) fitting photometry
package (\markcite{pbs94}Stetson 1994).  The interactivity of this
package enabled us to select $\sim$170 PSF stars well-distributed
spatially over the frame.  This was critical in accurately
characterizing the significant radial variation in PSF across the
frame.  For each frame, the PSF was iteratively improved by fitting
and removing faint neighbor stars.  The final PSF was applied to the
image using ALLSTAR.  A second application of ALLSTAR yielded a frame
with all the stars subtracted except a set of $\sim$220 bright,
isolated stars.  This frame was used to derive aperture corrections by
subtracting the PSF-fit DAOPHOT magnitudes of the stars from their
aperture magnitudes (derived using the same software and parameters as
those used in measuring the standard stars).  The aperture correction
was then applied to all the stars measured in the frame, and the
standard star relations were applied, yielding standard $VI$
magnitudes for thousands of stars on the on- and off-cluster fields.
The on-cluster $VI$ data provides a set of secondary standards for
calibrating the other on-cluster images.
 
The remaining images of NGC 6441 were reduced using a version of the
DoPHOT PSF-fitting package (\markcite{sms93}Schechter, Mateo, \& Saha
1993) which allows for spatial variations in the PSF.  Instrumental
magnitudes were obtained for each frame and combined into instrumental
$VI$ pairs, and then transformed to standard $VI$ magnitudes via fits
to the secondary standards described above.  We grouped the $VI$ frame
pairs as follows: good-seeing frames with ($a$) long, and ($b$) short
exposure times, and poor-seeing frames with ($c$) long, and ($d$)
short exposure times.  This grouping strategy minimized mis-matches
due to variable depth and seeing.  Within each group, we spatially
matched stars from the different frames, and combined the photometry
for each star using an error-weighted mean.  For the error in each
combined stellar magnitude, we adopted the standard error of the mean
for the individual measures, which proved to be a more conservative
estimate than computing the error directly from the DoPHOT errors.

\section{Color-Magnitude Diagrams}

We combined the mean photometry from groups $a$ and $b$ (the long- and
short-exposure frames with good seeing) using an error-weighted mean
to produce a single dataset with maximum dynamic range.  Outside a
radius of 1100 pix (7.33 arcmin) from the cluster center, elongated
star images sometimes registered as multiple stars, compromising the
mean photometry and complicating variable star detection (see \S 4).
Figure 2a shows the ($V-I$, $V$) color-magnitude diagram (CMD) for
22,871 stars on the on-cluster frame lying within 1100 pix of the
cluster center.  Only stars detected on four or more frames, and
having combined errors in $V$ and $V-I$ of $\sigma_V < 0.050$ mag and
$\sigma_{V-I} < 0.071$ mag, are shown.  Table 1 lists the combined
photometry for all the bright stars ($V < 19$ mag) within 1100 pix of
the cluster center.
 

 
Figure 2b shows the corresponding CMD for the off-cluster field.  This
diagram represents data from a single $VI$ frame pair, so the errors
are larger.  Poor photometry, galaxies, and cosmic rays were rejected
by performing 3-$\sigma$ cuts on the DAOPHOT values of ``$\chi$'' and
``sharp'' as functions of magnitude.  Only stars with $\sigma_V <
0.050$ mag and $\sigma_{V-I} < 0.071$ mag are shown (DAOPHOT errors).
 
Several features are apparent in these CMDs.  First, both the cluster
and field have large numbers of extremely red giant stars.  Second, in
both CMDs one can detect the red clump stars of the bulge at $V
\approx 16.8$ and $V-I \approx 1.7$ mag.  This agrees well with
photometry of the nearby field MM-5B from the OGLE group
(\markcite{smu94}Stanek et al. 1994).  Note that the red clump in
Figure 2b is shifted redward and faintward relative to that in Figure
2a, suggesting a mean reddening difference between the two fields.
Third, a prominent feature of the on-cluster CMD is the NGC 6441 red
clump, which forms a swath of stars near $V \approx 17.5$ and $V-I
\approx 1.4$ mag, running parallel to the reddening vector (we adopt
$A_V = 2.6~E(V-I)$ throughout this paper).  Finally, the on-cluster
CMD shows a population of stars bluer and somewhat brighter than the
cluster red clump.  Such stars are not present on the off-cluster CMD,
and represent the first ground-based view of the extended blue
horizontal branch seen by \markcite{rmr97}Rich et al. (1997).
 
Figures 3a and 3b zoom in on the HB regions of these CMDs.  The tight
distribution of red clump stars in Figure 3a paralleling the
reddening vector suggests that there is significant differential
reddening across the face of the cluster.  This also explains why a
tight red giant branch (RGB) is not seen; differential reddening
spreads the RGB stars roughly perpendicular to the un-reddened RGB
fiducial. We will consider the reddening problem in more detail in \S
5.2.
 


Another important feature is highlighted in Figure 3c, where stars
from Figure 3a lying between 40--160 arcsec of the cluster center are
displayed.  The excess of blue HB (BHB) star candidates seen in Figure
2a is evident, indicating that they too are members of the cluster.
The cluster's red clump and red giant branch are better defined in
this panel as well.
 
Figure 3d shows another effort to isolate NGC 6441 stars from the
field.  Here, we have statistically subtracted the CMD of off-cluster
stars located between 40--160 arcsec of the frame center from the
stars in Figure 3c.  Before subtracting, we corrected for a 0.11 mag
reddening difference between the fields (the off-cluster field has the
higher reddening), which we estimated from the shift in color
histograms of stars having values of the reddening-independent
parameter $V_{V-I} = V - 2.6~(V-I)$ between 12.1 and 13.3
(\markcite{smu94}Stanek et al.  1994).  For each star in the
off-cluster CMD, we removed from the on-cluster CMD the star nearest
in color-magnitude space.  Though the subtraction is over-aggressive
for the faint stars, the brighter sequences (including the blue HB,
red clump, and red giant branch) are more clearly shown.

The lack of faint, extremely blue HB stars in Figure 3c, relative to
their abundance in the CMD of \markcite{rmr97}Rich et al. (1997),
comes as a mild surprise.  Even in CMDs where we relax the criteria on
the photometric errors, these stars do not appear at $V > 18$ mag.
This suggests that these stars may be radially concentrated toward the
cluster center, but we can not state this unequivocally without
detailed completeness modeling.  A simpler, more direct comparison
with the \markcite{rmr97}Rich et al. (1997) $BV$ data will soon be
available from the deep, ground-based $BV$ photometry of the cluster
recently obtained by Horace Smith et al. (private communication).

Does the prominence of the less extreme BHB (i.e., brighter than $V
\approx 18$ mag) vary with radial distance from the cluster center?
\markcite{rmr97}Rich et al. (1997) found no evidence for a change in
the number ratio of BHB to red clump stars with radius over their
small {\em HST} field of view ($\sim$100 arcsec, with the PC chip
centered on the cluster), and interpreted this as evidence against the
production of BHB by dynamical effects.  We can extend this search to
larger radii using our $VI$ photometry.  For this purpose, we defined
BHB stars as having $16.8 < V < 18.0$ and $0.3 < V-I < 0.8$ mag, and
red clump stars as having $13.6 < V_{V-I} < 13.9$ and $1.30 < V-I <
1.56$ mag.  We defined the number fraction
\[ F = (N_{BHB,c} - N_{BHB,f}) / (N_{RC,c} - N_{RC,f})\]
where the $c$ and $f$ subscripts indicate stars counted from the
cluster and field photometry sets, respectively.  For stars in the
radial annulus 50--100 arcsec, we obtained $F = 0.054 \pm 0.012$,
while for radii of 100--240 arcsec, we obtained $F = 0.027 \pm 0.008$
\footnote{The star counts are: $N_{BHB,c} = 23$, $N_{BHB,f}= 1$,
$N_{RC,c} = 417$, $N_{RC,f}= 11$ for the inner annulus.  For the outer
annulus, they are 21, 7, 598, and 77, respectively.}.  The blue HB
stars are a factor of two more prominent in the inner annulus.

The Fokker-Planck simulations of \markcite{rmr97}Rich et al. (1997)
indicated that BHB stars produced from dynamically-stripped RGB stars
should be concentrated within 4 core radii (26 arcsec,
\markcite{wmh96}Harris 1996).  The lack of BHB concentration in the
\markcite{rmr97}Rich et al. (1997) observations, and our observed
concentration at larger radii do not appear to be consistent with
simple pictures of BHB star production through tidal stripping of
normal RGB stars by close encounters in the dense cluster core.

In many clusters, relative radial population gradients are attributed
to the effect of mass segregation.  The relaxation time for NGC 6441
at the half-mass radius, $\sim$0.5 Gyr at 38 arcsec
(\markcite{wmh96}Harris 1996), is much longer than the typical
lifetime of an HB star ($\sim$0.1 Gyr), but mass segregation could be
at play if the BHB stars are members of binary systems.
\markcite{bai95}Bailyn (1995) reviewed two scenarios for producing
extreme BHB stars from binaries: the merger of helium white dwarf
binaries, and mass transfer in wide binaries in which the hydrogen
envelope of a red giant evolving close to the RGB tip is donated to
the secondary star.  \markcite{bea92}Bailyn et al. (1992) found the
latter scenario to be particularly appealing in explaining the radial
distribution of extreme BHB stars in the globular $\omega$ Centauri,
and the scenario may be relevant to NGC 6441 as well.  It explains the
radial concentration of BHB far from the cluster center, and the broad
color distribution of the BHB population (the latter due to a range in
mass loss, produced by a range in orbital separations).  On the
negative side, the scenario does require NGC 6441 to have a much
larger population of binaries with the appropriate orbital periods (a
few hundred days) than metal-rich clusters with ``normal'' red-clump
HBs.  Observationally, there are few constraints on the distribution
of binary periods in globular clusters.  Theoretically, the primordial
period distribution of a cluster's binary stars could be linked to the
angular momentum of the gas cloud which formed the cluster.  Also,
dynamical evolution of the binary populations will vary from cluster
to cluster (e.g., \markcite{bea92}Bailyn et al. 1992).  Thus there
seems to be no argument against the existence of large variations in
the binary period distribution from one cluster to another.  Finally,
the wide-binary scenario makes a prediction: most BHB stars should be
members of a binary system, with radial velocity variations of a few
tens of km~s$^{-1}$.  Such observations should be obtainable using
multi-fiber spectrographs on the new generation of large telescopes.

\section{Variable Stars}

We used the \markcite{wns93}Welch \& Stetson (1993) variability index,
$I_{WS}$, to characterize the likelihood that a given star is a
variable.  The value of $I_{WS}$ is computed for each star from the
individual $V$ and $I$ magnitudes and their errors.  For a
non-variable star, $I_{WS}$ approaches zero as the number of
observations becomes large, whereas for a variable, $I_{WS}$
approaches a constant value.  To minimize the effects of seeing and
exposure depth (i.e., the blending of two nearby stars into a single
measurable image as the seeing and/or signal-to-noise conditions
worsen), we employed the frame groupings described in \S 2 and
computed $I_{WS}$ for each star detected more than four times in each
group.  Groups $(a)$ and $(b)$ were useful in detecting variable stars
near the cluster ($R < 1100$ pix), and groups $(c)$ and $(d)$ were
useful outside $R \approx 600$ pix\footnote{An alternative approach
would have been to develop a master list of star positions from the
best seeing frames, and to measure magnitudes for all the stars in
this list on each frame.  In practice, the residuals of the fits used
to transform the stellar coordinates from one frame to the next were
large relative to the tolerance required for accurate PSF-fitting
using fixed coordinates.  However, the residuals were sufficiently
small to enable reliable matching of stars from frame to frame.}.
 
The morphology of the points in the ($I_{WS}$, $V$) plane led us to
classify stars as probable variables ($I_{WS} > 170$) and possible
variables ($50 < I_{WS} < 170$).  When plotted on the cluster CMD, it
became clear that three classes of variables are present: extremely
red variables (Mira, semi-regular, and irregular variables, hereafter
LPVs), blue variables (possible RR Lyrae stars), and a few other
variables scattered around the CMD.  We focused on these stars as
follows: all stars with $V-I > 2.4$ mag and $I_{WS} > 50$ (LPV
candidates), all stars with $V-I < 1.4$ mag and $I_{WS} > 50$ (RRL
candidates), and all other stars with $I_{WS} > 120$.  We enhanced our
search for cluster RRL stars by supplementing the RRL candidates with
stars in the subset $V-I < 1.5$ mag, $16.8 < V < 17.8$, and $30 <
I_{WS} < 50$.  We produced analogous lists from the data in groups
$(b)$, $(c)$, and $(d)$, and merged them into a final list of 115
variable star candidates.
 
We then plotted the $V$ and $I$ magnitudes as a function of time for
each candidate.  The LPV candidates were characterized by constant or
slowly-varying $V$ and $I$ magnitudes, with distinct jumps between the
May and June data.  Two candidates near the cluster center were
sufficiently crowded, and their light curves scattered, that we could
not unambiguously characterize them as LPVs; their positions and
magnitudes are recorded in Table 5 (stars SV6 and SV7).  The remaining
LPV stars are considered in \S 4.2.
 
Among the 56 RRL candidates, 33 had time-magnitude plots which
suggested that a close pair of non-variable stars in some frames had
been resolved and in others was detected as a single star; when the
photometry from the various frames was matched (based on position),
the blended photometry was systematically brighter than the resolved
photometry, and a large value of $I_{WS}$ resulted.  Inspection of the
images almost always confirmed the blended-neighbor hypothesis.  We
discuss the remaining 23 candidates in detail in \S 4.3.
 
Of the 16 variable star candidates scattered over the CMD, all but one
were found to be false detections of critically-resolved, non-variable
pairs.  The single true variable has a color ($V- I \approx 2.1$)
consistent with an LPV near maximum light, so we have grouped it with
the other LPV stars in Table 2 (V35).  Clearly, the large number of
critically-resolved stars in this dense stellar field presents a
complication to the identification of variable stars, but one that is
surmountable given careful attention.

\subsection{Previous Studies of Stellar Variability}
 
The {\em Catalogue of Variable Stars in Globular Clusters}
(\markcite{hsh73}Hogg 1973), as updated by \markcite{cc_97}Clement
(1997), lists twelve variable stars in NGC 6441.  No photometric or
variable-type information is given.  Stars V7 and V8 lie outside our
field of view.  Only variables V1, V6, and V9 were detected as
variables in our study.  Examination of the remaining stars'
photometry showed that V2, V3, V5, and V10 are all LPVs, but were
absent from enough frames due to saturation and other effects to
sidestep detection.  V4 showed no sign of variability in our
observations, but at $V-I = 2.84$, is red enough to be an LPV (see \S
4.2).  V11 and V12 are much fainter, bluer stars which
\markcite{hnh76}Hesser \& Hartwick (1976) suspected as being RR Lyrae
stars; our photometry shows no sign of variability in these stars (see
Table 5 and the Appendix).

\subsection{Long-Period Variable Stars}
 
Data for the 31 LPV stars (i.e., stars which varied on timescales
longer than a few days) found near NGC 6441 are presented in Table 2.
The first column gives the variable star designation (in roughly
decreasing value of the $I_{WS}$ variability index).  We name our
newly-discovered variables V13, V14, etc.  The second and third
columns give the position of the star in the $X$ and $Y$ coordinate
frame of Figure 1 (in units of pixels).  The right ascension and
declination offsets from the cluster center (in arc seconds, with
positive values indicating East and North) can be computed from the
pixel positions as follows,
\[ \Delta \alpha = -0.396 (X_{\rm pix}-1014),~~~~~ \Delta \delta = -0.396 (Y_{\rm pix}-1000) \]
The remaining columns give the magnitude-mean $I$ magnitude
(${\overline{I}}$), the mean color (${\overline{V-I}}$), and the
number of observations ($N$) during the May and June observing runs.
In $\sim$40\% of the stars, we observed a significant change in
brightness or color during one or both observing runs.  The previously
identified LPVs not detected in our study (see \S 4.1) are included at
the bottom of this table.
 

Clearly, the data here do not fully sample the LPVs' light cycles, and
thus do not accurately represent the stars' mean photometric
properties.  Still, the locations of the stars on the CMD, shown in
Figure 4, approximately represent these mean positions.  Furthermore,
the limited phase coverage leads to LPV detection incompleteness; the
observed $V$ and $I$ magnitude differences of some true LPVs (e.g.,
V4) were insufficient to result in $I_{WS}$ values large enough to
trigger detections.  Finally, saturation and other image complications
left some stars with too few observations to fulfill all the
variability detection criteria.  Thus, it is likely that many more of
the stars with $V-I > 2.4$ are LPVs, in particular faint, red LPVs
near the cluster center.



\subsection{RR Lyrae Variables and Eclipsing Binaries}

In an effort to determine the types of variable stars present among
the 23 blue variable star candidates discussed above, we attempted to
find the period of each star and create the star's mean light curve.
Classical period finding techniques such as Fourier transforms or
Phase Dispersion Minimization perform poorly, if at all, when the data
are sparse and irregularly spaced.  We therefore developed the
following new method.

In cases where a star's variable type is suspected (e.g., an RR
Lyrae), it is possible to assume a period and fit the folded light
curve with an appropriate template light curve.  If the assumed period
is near the true period, the scatter about the fitted template will be
small.  By folding the time-magnitude data by a sequence of periods
typical of the variable type, and fitting the template and computing
the scatter (or $\chi^2$) at each period, one can easily identify
$\chi^2$ minima indicative of possible periods.  This method is
similar to that described by \markcite{pbs96}Stetson (1996).

In practice, we used the template-fitting procedure described in
\markcite{acl98}Layden (1998).  We fit each star with three templates;
$V_3$ and $V_4$ from \markcite{acl98}Layden (1998, Table 4) are
typical of RRa and RRb type variables, respectively, while a cosine
curve was used to represent RRc and contact binary LCs.  We fit the
templates over periods ranging from 0.2 to 1.0 days, increasing the
period by 0.1\% from one step to the next.  If no suitable minima were
found, we searched on the period interval 1--3 days.  Periods longer
than 3 days are evident in the time-magnitude plots.  Since the
templates are seldom a perfect match for the actual LCs, we refined
the periods by folding the time-magnitude data by a sequence of
periods near the $\chi^2$ minimum, plotting the resulting LCs, and
chosing the cleanest LC.  This resulted in tighter LCs and improved
periods.

We note that this technique should be applicable to any type of
variable with a characteristic LC.  Indeed, by including template LCs
from a large number of variable types, it should be possible to
determine both period {\em and} variable star type from the template
giving the smallest $\chi^2$ value.

Of the 23 blue variable candidates, we were able to determine periods
for 11 RR Lyrae stars (V36--V46) and two contact binaries (V48 and
V50).  Light curves for two detached binaries (V49 and V51) and one
long period star (V6, a Cepheid?) were obtained by examination of the
time-magnitude plots.  The $V$ and $I$ light curves of these 15 stars
are presented in Figure 5.  Possible periods were obtained for five
additional stars (SV1--SV5, see Table 5).  We were unable to determine
with certainty whether these stars were low-amplitude RR Lyrae or
contact eclipsing binaries (see Figure 6).  We were unable to obtain
periods or LCs for the remaining two blue variable candidates, which
we list in Table 5 as SV8 and SV9.





Table 3 lists the photometric properties of the 11 RR Lyrae stars.
Columns include the location of the stars in the coordinate frame of
Figure 1 ($X_{\rm pix}$ and $Y_{\rm pix}$), the period in days, the
intensity-mean $V$ and $I$ magnitudes ($\langle V \rangle$ and
$\langle I \rangle$), the $V$-band amplitude ($\Delta V$), the mean
color during the phase interval 0.5--0.8 ($(V-I)_{\rm min}$), and a
brief comment.  For several stars, a few discrepant data points were
omitted from the computations, (e.g., V37), while for several other
stars, only good-seeing data were used (e.g., V42 and V43).  These
data will prove useful in determining the nature of these unusual
stars in the following section.


Table 4 reports the positions of the four eclipsing binaries from
Figure 5, along with their periods, $V$ and $I$ band maxima and
amplitudes, and a comment.  Table 5 lists the positions,
magnitude-mean $V$ magnitudes and $V-I$ colors of the uncertain and
suspected variables, along with a brief comment.


The time series photometry for each of the variables is given in Table
6.  The columns include the variable star name, the heliocentric
Julian Date of mid-observation ({\it HJD}), the $V$-band magnitude and
its error ($\sigma_V$), the $I$-band magnitude and its error
($\sigma_I$), and a quality code ($Q$) which corresponds to the
symbols used in the light curves: $Q=1$ indicates a long exposure
under good seeing conditions ($\bullet$), $Q=2$ indicates a short
exposure in good seeing ($\circ$), $Q=3$ indicates a long exposure in
poor seeing ({\em solid triangle}), and $Q=4$ indicates a short exposure in
poor seeing ($\bigtriangleup$).  Figure 7 shows the finder charts for
each variable star.




\subsection{Cluster Membership}

The positions of the variable stars in the CMD and their radial
locations with respect to the centroid of the cluster light provide
useful information regarding the stars' membership in the cluster.
Figure 4 shows the variable stars placed in the CMD, while Figure 8
shows the cumulative radial distributions of the stars in various
classes.  The squares in Figure 8 give the total sky-subtracted
$V$-band photon counts inside the radius $R$, normalized by the amount
at $R=440$ arcsec.  The crosses give the analogous counts expected if
a uniform stellar distribution were spread across the usable pixels of
the CCD.  These curves represent the {\em approximate} distributions
expected for cluster member and field stars, respectively.  The {\em
actual} observed distribution for a given class of star will deviate
from these expectations due to brightness and crowding effects.  The
lines in Figure 8 are the normalized cumulative radial distributions
of various classes of stars, selected from the CMD as follows.  Red
clump stars ({\em solid line}) lie in the parallelogram bounded by
$13.6 < V_{V-I} < 13.9$ and $1.3 < V-I < 1.56$, blue HB stars ({\em
dotted line}) lie in the rectangle $16.8 < V < 17.6$ and $0.3 < V-I <
0.9$, while the LPV ({\em long dash}), RR Lyrae ({\em short dash}),
and eclipsing binaries ({\em dash-dot line}) were taken from Tables 2,
3, and 4, respectively.  The radial locations of the suspected
variables in Table 5 are indicated along the top of the figure.



First, consider the four well-established eclipsing variables (Table
4).  These stars are scattered about the CMD (Figure 4) and their
location relative to the cluster is consistent with a uniform
distribution across the CCD field (Figure 8).  Thus, it is reasonable
to assume that these stars are members of the foreground stellar
populations, probably the old disk (\markcite{ruc97}Rucinski 1997).

Next, consider the RR Lyrae stars from Table 3.  In the CMD, eight of
these stars are clustered near $(V-I, V)$ = (1.1, 17.3), just blueward
of the prominent cluster red clump, and just redward of the blue HB
stars. The radial distributions of the RR Lyrae, red clump, and blue
HB stars all resemble that of the cluster light (Figure 8).  Of these
three classes of stars, the red clump stars are the least radially
concentrated, as expected if a significant number of field red giants
contaminate the counting box.  The BHB stars are more centrally
concentrated, and the RR Lyrae stars seem even more so, though the
number of stars involved is small and the expected background
contamination varies between the three stellar classes.  Note that all
three of these distributions drop precipitously within 1 arcmin of the
cluster center, indicating that our detection completeness decreases
at smaller radii, reaching zero inside $\sim$30 arcsec.  It is highly
probable that these eight RR Lyrae belong to the cluster.

Of the other three RR Lyrae from Table 3, V36 is almost certainly a
foreground star.  Its location 297 arcsec from the cluster, its
position in the CMD just blueward of the bulge red clump, and its
light curve parameters (see \S 5.1) are all consistent with it being a
member of the Galactic bulge.  The nature of V41 and V44 is less
clear, and discussion of these stars is reserved for \S 5.3.

The LPV stars from Table 2 have a distribution in Figure 8 very
similar to that of the cluster light between 1 and 4 arcmin.  Inside 1
arcmin, incompleteness sets in, though not as quickly as for the
fainter HB stars.  Suspected variables SV6 and SV7 are probably
cluster LPVs whose photometry has been compromised by stellar
crowding.  Thus, most of the LPVs in Table 2 probably belong to the
cluster.  In the CMD, most of the LPVs form a fairly tight, linear
sequence running from $(V-I, V)$ = (2.6, 15.6) to (5.0, 18.0).  A set
of five LPVs appears to parallel this sequence $\sim$0.8 mag brighter.
These five stars (V3, V5, V9, V30, and V31) have a cumulative radial
distribution consistent with a uniform foreground population.  Given
that the distance modulus difference between the cluster and the bulge
is $\sim$0.8 mag, based on the offset in $V$ magnitudes of the red
clumps and RRL stars, it seems reasonable to conclude that these five
bright LPVs belong to the bulge, rather than the cluster.

The population of the two LPVs $\sim$1.0 mag fainter than the cluster
LPV sequence is unclear.  Their magnitudes and large radial distances
from the cluster ($R = 215$ arcsec for V25 and 322 arcsec for V28)
suggest they are members of the background field.  It is conceivable
that they belong to the Sagittarius dwarf galaxy, but ($i$) the
distance modulus between Sagittarius and the bulge is $\sim$2.5 mag
(\markcite{snl95}Sarajedini \& Layden, 1995), 0.7 mag too large for
the observed offset, and ($ii$) NGC 6441 is over 10 degrees from the
nearest part of Sagittarius in the map of \markcite{igi97}Ibata et
al. (1997).  Complete light curve coverage of these variables would
enable estimates of their distances, while spectroscopy would indicate
whether their radial velocities are consistent with an origin in the
Galaxy or in Sagittarius.

Finally, we consider the variables of uncertain type from Table 5.  Of
these five stars, SV2, SV3, and SV5 lie in the clump of bona-fide
cluster RRL in Figure 4, and so we tentatively consider them to be RRL
as well.  Their radial locations (Figure 8) are consistent with those
of the other RRL.  While SV4 has the correct $V$ magnitude, it is 0.3
mag bluer than the other RRL.  Whether SV1 is a c-type RR Lyrae or a
contact binary, its location in the CMD clearly indicates that it
belongs to the foreground field.

\section{Discussion}

\subsection{The Period-Amplitude Diagram}

The Period-Amplitude diagram is a useful, reddening-independent tool
for comparing RR Lyrae variables.  For typical RRL in the field and in
metal-poor globular clusters, a star's position in this plane depends
on its metallicity.  In Figure 9, we show the metal-rich field ab-type
RRL (periods and amplitudes from \markcite{bmy77}Bookmeyer et
al. 1977, metallicities from \markcite{acl94}Layden 1994).  At a given
amplitude, the more metal-rich stars have systematically shorter
periods.  The ab-type RRL in the globular cluster NGC 6171
(\markcite{dic70}Dickens 1970; $[Fe/H] = -1.04$,
\markcite{wmh96}Harris 1996) overlie the field RRL of similar
metallicity, while that cluster's c-type RRL congregate near ($\log
P$, $\Delta V$) = (--0.5, 0.5).

The RR Lyrae stars in NGC 6441 are systematically displaced to longer
periods and smaller amplitudes.  The bulge RRL V36 is marked by the
large asterisk at $\Delta V \approx 1.2$, a position suggesting a
metallicity of $[Fe/H] = -1.0$ to -1.2, consistent with the
distribution of metallicities in bulge RRL (\markcite{wnt91}Walker \&
Terndrup 1991).  The large asterisks at lower amplitude mark the
anomalously red variables V41 and V42.  The open circles mark the
suspected c-type RRL SV2, SV3, and SV5.  The solid circles mark the
remaining RRL in Table 3.

These results are consistent with the positions of the three
previously-known long-period, metal-rich RRL: V9 in 47 Tuc
(\markcite{csw93}Carney et al. 1993), and V28 (ab-type) and V27
(c-type) in NGC 6388 (\markcite{ssb94}Silbermann et al. 1994).
However, our data have more than tripled the number of members of this
interesting sub-class of variable.  Also, the positions of SV3, SV5,
and perhaps SV2 support our interpretation that these stars are c-type
RRL in NGC 6441.


\subsection{Reddening}

Using a variety of methods, \markcite{hnh76}Hesser \& Hartwick (1976)
estimated the interstellar reddening toward NGC 6441 to be $E(B-V) =
0.46 \pm 0.15$ mag.  They noted that, based on the appearance of the
Palomar Sky Survey prints, ``...the obscuration in the immediate
vicinity of NGC 6441 is approximately uniform...''.  However, the
large uncertainty they quote reflects the large range of solutions
produced by the different techniques, $0.23 \leq E(B-V) \leq 0.7$.
Furthermore, the large angular extent over which some of their
estimates were obtained may produce deceptive results if significant
patchiness exists on scales of several arcminutes or less.
\markcite{rjz80}Zinn (1980) and \markcite{rea88}Reed et al. (1988)
obtained $E(B-V) = 0.47$ and 0.49, respectively, from integrated light
measurements.

The RR Lyrae variables we have detected toward NGC 6441 provide an
excellent opportunity to determine the reddening in a small region
around the cluster.  \markcite{mus95}Mateo et al. (1995a) showed that
the de-reddened $V-I$ colors of typical RR Lyrae near minimum light
(phases 0.5--0.8) is roughly constant.  We determine $(V-I)_0 = 0.57
\pm 0.02$ for the eight stars with $[Fe/H] > -1.3$ listed by
\markcite{mus95}Mateo et al. (1995a).  We use the apparent
minimum-light colors of the 11 RRL in Table 3, $(V-I)_{\rm min}$, to
determine the reddening, $E(V-I)$ along each line of sight (see Table
7).  These values are converted into $E(B-V)$ and $A_V$ using the
relations of \markcite{dea78}Dean et al. (1978).


Of these stars, only V36 is a ``typical'' RRL, judging from its
position in the Period-Amplitude diagram and its expected location in
the Galactic bulge.  Indeed, one could question whether the
$(V-I)_{0}$ colors of bulge RRL are the same as the local stars used
to calibrate the relation.  It is reassuring that V36 gives $E(B-V) =
0.46 \pm 0.03$ mag, in excellent agreement with the existing reddening
values.

Does the relationship work for the atypical, long-period variables in
NGC 6441?  Unfortunately, $V-I$ colors do not exist for the two other
known members of this class, V9 in 47 Tuc and V28 in NGC 6388.
However, the data from \markcite{csw93}Carney et al. (1993) on V9
indicates that the analogous relation employing $(B-V)_{\rm min}$,
period, and metallicity (\markcite{bla92}Blanco 1992 and references
therein) {\em does} work, yielding $E(B-V) = 0.03$.  For comparison,
\markcite{wmh96}Harris's (1996) compilation lists $E(B-V) = 0.05$.
With this encouragement, and excluding the two anomalously red and
bright stars V41 and V44, we find the mean $E(B-V)$ of the nine RRL
toward NGC 6441 to be 0.45 mag (0.05 mag rms).  Again, our reddening
estimate is in excellent agreement with the existing estimates.

The rms value is somewhat larger than expected from the estimated
uncertainties ($\sim$0.03 mag for a single star).  This suggests there
may be some differential reddening across the face of the cluster.
The extension of the cluster red clump along the reddening vector and
the absence of a well-defined red giant branch provide similar
evidence (see \S 3).

To search for further evidence of spatially-variable reddening, we
considered individually the CMDs of stars lying in eight equal-sized
sectors of an annulus with radius 40--100 arcsec centered on the
cluster (the angles $\alpha_1$ and $\alpha_2$ in Table 7 define the
bounds of each sector, and are measured counter-clockwise from a line
radiating in the +$x$ direction from the cluster center in Figure 1).
For each CMD, we isolated the red clump stars using $13.6 < V_{V-I} <
13.9$ and produced their generalized color histogram (see Figure 10).
Several curves are substantially displaced towards redder colors,
indicating the presence of differential reddening.  Sector 4 is
particularly noticeable.  The median $V-I$ color of each histogram
($Med(V-I)^{13.6}_{13.9}$), and the number of stars involved ($N_{\rm
sec}$), are shown in Table 7.  The widths of the individual histograms
are not drastically larger than the color width of the red clump in 47
Tuc ($\sim$0.2 mag, \markcite{tea88}Armandroff 1988), suggesting that
the effects of differential reddening within a given sector are small,
though red giant branches in the CMDs are still noticeably broadened
(especially in Sector 4).  The small-scale peaks in the Figure 10
histograms are due to shot noise, indicating that further spatial
segregation of the cluster is not warranted.



In an effort to recover the de-reddened CMD of the cluster, we
corrected the differential reddening in the individual sectors to a
representative mean ($\langle Med(V-I)^{13.6}_{13.9} \rangle = 1.420$,
the average of Sectors 1, 2, 5 and 6), and then corrected the entire
dataset for the average RRL reddening ($\langle E(V-I) \rangle =
0.577$, based on the RRL stars in or near those sectors).  We assumed
$A_V = 2.6~E(V-I)$.  The de-reddened CMD is shown in Figure 11.  The
principal sequences seem to have tightened up somewhat, compared with
Figure 3c.  However, the slope of the red clump stars persists,
confirming the \markcite{rmr97}Rich et al. (1997) {\em HST}
photometry.  \markcite{snc98}Sweigart \& Catelan (1998) note that this
intrinsic slope rules out older cluster age or extreme mass loss on
the giant branch as causes of the extended blue HB seen in NGC 6441.



\subsection{V41 and V44}

The nature of the variables V41 and V44 is unclear.  Their LCs suggest
they are RR Lyrae stars, or possibly another class of pulsating
variable.  They are both $\sim$0.7 mag brighter, and $\sim$0.3 mag
redder, than the other eight cluster RR Lyrae.  Both are located
close to the cluster center (63 and 57 arcsec, respectively), so are
likely to be members.

One possibility is that they are Type II Cepheids belonging to the
cluster.  However, we searched for periods between 1 and 5 days and
found no reasonable minima, and the time-magnitude plots indicate
shorter period variations.  Anomalous Cepheids have periods between
0.36 and 1.6 days, absolute magnitudes between 0.5 and 2 mag brighter
than RRL, and colors comparable to RRL (\markcite{nea94}Nemec et
al. 1994).  Based on the first two criteria, V41 and V44 could be
anomalous Cepheids.  However, both stars are significantly redder than
the RRL in NGC 6441.  Furthermore, most anomalous Cepheids are found
in dwarf spheroidal galaxies, metal-poor systems with a component of
younger stars ($\leq$5--10 Gyr, \markcite{mfk95}Mateo et al. 1995b).
\markcite{bc97a}Bono et al. (1997a) confirm theoretically that the
minimum mass for the formation of anomalous Cepheids is $\sim$1.8
$M_{\odot}$ at $[Fe/H] = -1.7$, and this minimum increases quickly with
increasing metal abundance.  It seems unlikely that NGC 6441 possesses
a population of stars young enough to produce HB stars of this mass.
It may also be possible to produce anomalous Cepheids through binary
star coalescence, but at the metallicity of NGC 6441, the mass limit
appears incompatible with turnoff stars of globular cluster age ($M
\approx 0.8 M_{\odot}$).  Given these complications, and that we are
not aware of the existence of any metal-rich examples of this type of
variable, it seems unlikely that V41 and V44 are anomalous Cepheids.

Another possibility is that V41 and V44 are RR Lyrae stars in the
foreground bulge.  The similarity of their $V$ magnitudes to that of
V36 supports this idea, but their much redder $V-I$ colors would
require a large amount of differential reddening.  The two stars {\em
do} lie near each other, 21 arcsec apart in Sector 3; perhaps they lie
behind a small, dense foreground dust cloud.  To test this, we
extracted a subset of the stars in Table 1 which lie within 20 arcsec
of the stars' centroid, and plotted their CMD.  It shows no
displacement relative to the CMD of the other stars in Sector 3, nor
to Figure 3c, indicating that heavy differential obscuration is {\em
not} responsible for the anomalous colors of these stars.
Furthermore, the central location of V41 and V44 argues strongly for
cluster membership.  If they belong to the field, the probability of
both variables lying inside 63 arcsec radius is $4\times10^{-4}$.

A third possibility is that V41 and V44 are cluster RRL whose
photometry has been contaminated by another star.  If the RRL
components are typical of the other eight cluster RRL, then the
contaminating stars must have $V \approx 17.4$ and $V-I \approx 1.6$,
roughly like those of cluster red clump stars.  The positions of V41
and V44 in the Period-Amplitude diagram (Figure 9) are consistent with
this interpretation.  The observed amplitudes are small compared with
the other RRL in NGC 6441 with the same period, but the amplitudes
become consistent with these stars when corrected for the effect of
the photometric companion just described: the $V$-band amplitude of
V41 becomes $\sim$0.7 mag, and of V44 becomes $\sim$1.2 mag.  Finally,
the values of the ``sharp'' parameters produced by DAOPHOT are typical
of stars of their brightness, indicating that the angular separations
of the contaminating stars are small compared to the seeing disk.
Perhaps V41 and V44 are members of physical binaries with red clump
stars.  This would be compatible with the wide-binary hypothesis for
BHB star formation discussed in \S 3.  In the case of V41 and V44, the
mass-enhanced secondary has evolved onto the red HB, whereas for the
other RRL and BHB stars, the secondary is still on the main sequence,
and contributes little to the combined light ($\lesssim 0.1$ mag).

We conclude that V41 and V44 are probably RR Lyrae members of NGC
6441, but appear brighter and redder than the other cluster RRL
because their photometry has been contaminated by the light of
unresolved red clump companions.  Images with higher spatial
resolution and detailed spectroscopy may confirm this conclusion
(unfortunately, V41 and V44 were just off the {\em HST} images of
\markcite{rmr97}Rich et al. 1997).

\subsection{Light Curve Shapes}

The light curves of ``normal'' c-type RR Lyrae in clusters and the
field have a characteristic shape, roughly sinusoidal, but with the
region of maximum brightness skewed toward smaller phases.  That is,
the phase interval of ``rising light'' between minimum and maximum
brightness is shortened, $\Delta \Phi_{\rm r} < 0.5$.  Among the 31
c-type RRL in the {\em General Catalog of Variable Stars}
(\markcite{gcvs4}Kholopov 1985) with $V$-band photometry, $\Delta
\Phi_{\rm r}$ ranges between 0.32 and 0.46, with a mean of 0.39 and
rms of 0.04.

Figure 6 shows that the LC shapes of the c-type RRL in NGC 6441 depart
somewhat from this norm.  Their maxima appear skewed to longer
periods, and $\Delta \Phi_{\rm r} > \sim$0.5.  The minima may be
uncharacteristically sharp as well, especially in the case of SV3.
Clearly, better LCs (more observations with smaller errors) for more
stars are required to verify this phenomenon.  When the observations
are confronted with stellar pulsation models (e.g.,
\markcite{bc97b}Bono et al. 1997b), this effect may prove to be a
valuable tool for constraining the parameters responsible for bright,
blue HB extensions in NGC 6441 and NGC 6388 (e.g.,
\markcite{snc98}Sweigart \& Catelan 1998).  The light curve shapes of
the ab-type RRL in NGC 6441 appear to fall within the range of shapes
exhibited by ``normal'' ab-type RRL.

\subsection{Comparison with Extended BHB Models}

\markcite{rmr97}Rich et al. (1997) concluded that neither age nor
dynamical effects, individually, could account for the extended blue
HB in NGC 6441, though they could not rule out a combination of the
two.  \markcite{snc98}Sweigart \& Catelan (1998) used synthetic HB
models to explore the effects of changing parameters which may produce
blue HB extensions.  They concur that neither age nor enhanced RGB
mass loss can produce the sloped HB, and hence luminous BHB extension,
seen in NGC 6441.  They considered three additional scenarios, (1)
high initial helium abundance, (2) core rotation, which at the RGB tip
allows both enhanced mass loss and formation of an over-massive core,
and (3) envelope helium abundance enhancement from deep mixing during
the RGB phase.

In their brief report, \markcite{snc98}Sweigart \& Catelan (1998)
could not present detailed results of their models.  We employ three
means of comparing their models and $BV$ synthetic CMDs with our $VI$
photometry: (a) the general distribution of stars in color and
magnitude on the HB, (b) the number ratio of HB stars to RGB stars
brighter than the mean RRL luminosity, $R$, and (c) the pulsation
characteristics of the RRL.

First, we compare the overall appearance of our observed CMD with the
synthetic CMDs of \markcite{snc98}Sweigart \& Catelan (1998, their
Figure 2) for their various scenarios.  The steep slope of our red
clump and its relation to the gentler-sloping BHB are best matched by
the scenarios with core rotation and helium-mixing, and moderate mass
loss on the RGB (see their Figures 2c and 2e).  Given the
uncertainties introduced by differential reddening, foreground
confusion, and by comparing the observed $VI$ CMD with the synthetic
$BV$ CMD, we can not make a more quantitative statement at this time.

Measuring the $R$-ratio in our photometry is complicated by the high
degree of contamination by stars in the foreground and by stars on the
cluster asymptotic giant branch (AGB).  Using as a guide a
field-subtracted version of Figure 11, we defined regions in which to
count the number of BHB, red clump, and red giant stars.  These
regions are shown in Figure 11.  We counted the number of stars in
each region on the unsubtracted, de-reddened cluster CMD (Figure 11)
and the equivalent de-reddened field CMD.  The counts are given in
Table 8.  We add nine RRL to the BHB counts and ten LPV stars to the
RGB counts.  We used the luminosity functions of
\markcite{ber94}Bertelli et al. (1994; $Z = 0.008$, $Y = 0.25$, Age =
14 Gyr) to estimate that 64\% of the stars in the RGB region are first
ascent red giants (36\% are AGB stars).  Combining the BHB and red
clump counts, we obtain $R = N_{\rm HB} / N_{\rm RGB} = 1.60 \pm
0.14$, where the error is due solely to Poisson statistics.
Systematic errors incurred by inappropriately defined counting regions
are surely higher.  If the counts in each region are uncertain by
25\%, the systematic error on our estimate of $R$ is 0.66.  Even with
this level of uncertainty, our counts strongly disfavor the high
initial helium abundance scenarios of \markcite{snc98}Sweigart \&
Catelan (1998).  Their simulations with main sequence helium
abundances of 0.38 and 0.43 produce sloping BHB extensions, but have
$R = 3.4$ and 3.9, respectively, about 3-$\sigma$ from our observed
value\footnote{In fact, we have underestimated the RGB counts in this
analysis.  The faint boundary of the RGB region is usually defined as
the bolometric magnitude equal to that of the RR Lyrae stars.
However, this limit passes through the red clump stars.  To avoid
miscounts, we raised the faint boundary to that shown in Figure 11.
Underestimating $N_{\rm RGB}$ results in overestimating $R$.  The true
value of $R$ is therefor less than 1.60, and the disagreement with the
model with high initial helium abundance is stronger.}.  Using the
relations of \markcite{bea83}Buzzoni et al. (1983), our value of $R$
implies a helium abundance of $Y = 0.25^{+0.05}_{-0.08}$, where the
errors are based on the larger, systematic errors quoted in $R$.
Using their $R^\prime$ parameter, we obtain $Y =
0.21^{+0.05}_{-0.07}$.  We caution that stellar interactions may alter
the relative numbers of HB and RGB stars, and thus affect the estimate
of $Y$.

The pulsation properties of the cluster RRL stars provide other
constraints on the \markcite{snc98}Sweigart \& Catelan (1998) models.
Their Figure 3 shows the positions of their model BHB stars (helium
mixing scenario with intermediate mass loss) in the temperature-period
diagram.  The dotted region in our Figure 9 encloses those models.  We
transposed the region from the ($\log T_{eff}$, $\log P$) plane to the
($\log P$, $\Delta V$) plane using Equation 1 of
\markcite{cat98}Catelan (1998), which relates $T_{eff}$ with $[Fe/H]$
and the $B$-band amplitude, $\Delta B$.  We converted from $B$ to $V$
amplitudes using the RRL in NGC 6171 (\markcite{dic70}Dickens 1970):
$\Delta V = 0.72~\Delta B + 0.03$ (rms = 0.04) mag, and we assumed
$[Fe/H] = -0.53$ dex.  Clearly, the model provides a reasonable,
though not perfect, match to the RRL stars in NGC 6441, NGC 6388, and
47 Tuc.  It will be interesting to compare the observations with the
temperature-period predictions of the other scenarios.

We can provide a measurement of the distribution of stars along the HB
which may prove useful in constraining future synthetic HB models, and
in comparing the HB of NGC 6441 with those of other globulars.  The
usual statistic is $(B - R) / (B + V + R)$, where $B$ and $R$ are the
numbers of HB stars on the blue and red sides of the instability
strip, and $V$ is the number of RRL variables.  We adopt the same CMD
regions as above except that the section of the BHB region redward of
$(V-I)_0 = 0.49$ (the mean de-reddened mag of the RRL in Table 7, see
counts in row labeled ``RHB'' of Table 8) is now added to the red clump
counts to produce $R$.  Subtracting the counts in the off-cluster CMD
from those in the on-cluster CMD, we obtain $B = 45 \pm 10$, $V = 9
\pm 3$, and $R = 648 \pm 29$ stars, so $(B - R) / (B + V + R) = -0.86
\pm 0.03$ (Poisson) $\pm 0.04$ (systematic).  The HB is predominantly
red, and only $\sim$17\% of HB stars are in the blue extension.


It is tempting to use our photometry to measure the distance to NGC
6441.  However, the long periods observed for its RRL suggest that
they are more luminous than typical RRL, so the absolute magnitudes
usually assumed for those stars may be inappropriate.  Moreover, the
different models presented by \markcite{snc98}Sweigart \& Catelan
(1998, their Figure 2) have absolute magnitudes near the instability
strip which range between 0.35 and 0.75 mag (we have omitted the high
initial helium scenarios).  The absolute magnitudes of the red clumps
present a smaller range of 0.80--0.95 mag.  Combined with a
de-reddened apparent red clump magnitude of $V_0 = 15.94 \pm 0.10$ mag
from Figure 11, we find that NGC 6441 lies between 11 and 13 kpc from
the Sun, on the far side of the Galactic bulge.


\section{Summary and Conclusions}

We present the first ground-based CCD photometry of the metal-rich
globular cluster NGC 6441.  Our $VI$ color-magnitude diagram shows
that the extended blue horizontal branch discovered by
\markcite{rmr97}Rich et al. (1997) using the {\em Hubble Space
Telescope} exists even at large radii from the cluster center.  Such
hot HB stars are not expected in a cluster of this metallicity.

Our time-series photometry enabled us to detect a large number of
variable stars.  We increased the number of red, long-period variables
from nine to 32.  We argue that five are members of the foreground
bulge and that two are members of the background field, or possibly
the Sagittarius dwarf galaxy.  The remaining 25 are members of NGC
6441.  We confirm the variability of V6, which may be a foreground
Cepheid.  Long-term monitoring of these variables is encouraged.

We also discovered 11 ab-type RR Lyrae variables, along with several
probable c-type RRL.  Based on their radial distances from the cluster
center and positions in the CMD, we argue that eight of the ab-type
RRL and 2--3 of the c-type RRL are members of NGC 6441.  One RRL
belongs to the foreground bulge, and two appear anomalously bright and
red.  We suspect the latter are photometric blends between cluster RRL
and red clump stars, and could be physical binaries.  We also
discovered four eclipsing binary stars in the foreground field, and
several stars which we could not classify with certainty.

We present a method for determining variable star periods in which the
data are folded by a sequence of periods, and at each period, are fit
with a set of light curve templates.  The $\chi^2$ of the fits are
then plotted as a function of the employed periods.  Minima in this
plot indicate probable periods.  

Using this method, we determined periods for the RRL, and produced
mean light curves in $V$ and $I$.  The 8--11 RRL in NGC 6441 have
periods that are systematically longer than field RRL of comparable
metallicity.  RRL in the metal-rich globulars 47 Tuc (V9) and NGC 6388
(V27 and V28) have similar properties.  Thus, a new sub-class of RRL
appears to be emerging.

Using the observed $V-I$ colors of the RRL at minimum light, we
determined the reddening toward NGC 6441 to be $E(B-V) = 0.45 \pm
0.03$ mag.  Two points suggest that these long-period RRL have the
same $(V-I)_0$ colors as the ``normal'' RRL, ($a$) we obtain $E(B-V) =
0.46$ mag from the bulge variable V36, and ($b$) the analogous
relation for $(B-V)_0$ at minimum light works for the long-period RRL
V9 in 47 Tuc.

Evidence for small changes in the amount of reddening across the face
of the cluster come from the star-to-star dispersion in the RRL
$E(B-V)$ values, and from differential shifts along the reddening
vector of CMDs generated from small regions around the cluster.  The
South-East portion of the cluster has $E(B-V) \approx 0.06$ mag higher
than the rest of the cluster, though there is evidence for patchy
reddening on a scale of several arcmin or less.

We employ a differentially de-reddened color-magnitude diagram to
count stars in various regions along the horizontal and red giant
branches.  We find the ratio of HB stars to red giants to be $R = 1.6
\pm 0.7$, indicating a normal helium abundance of $Y =
0.25^{+0.05}_{-0.08}$.  About 17\% of the HB stars lie blueward of the
red clump, and $(B - R) / (B + V + R) = -0.86 \pm 0.04$, indicating a
predominantly red HB.  We also find that the BHB stars are more
centrally concentrated than the cluster red clump stars.  This could
result from mass segregation if the BHB stars are members of binary
systems.  Furthermore, the BHB stars could be {\em produced} by mass
transfer in wide binary systems, as described by
\markcite{bai95}Bailyn (1995).  The anomalously bright and red RRL V41
and V44, which we have discovered in NGC 6441, could represent
binaries consisting of an RRL and a red clump star.  Spectroscopic
monitoring of these and the other BHB stars for binary motions would
provide stronger evidence for or against the binary hypothesis of BHB
star formation.

We also consider a different line of hypotheses, involving variations
in the physical properties of single stars, which might explain the
presence of the BHB stars' high surface temperatures.  As noted by
\markcite{snc98}Sweigart \& Catelan (1998), the intrinsic slope of the
HB, and in particular of the red clump, can not be reproduced by
standard models invoking extreme cluster age or red giant branch mass
loss.  Similarly, our measurement of the HB-to-giant star ratio, $R$,
argues strongly against the \markcite{snc98}Sweigart \& Catelan (1998)
model predictions for a scenario which involves high initial helium
abundance.  In contrast, the observed periods and amplitudes of our
RRL are in reasonable agreement with their model predictions for a
scenario in which the helium abundance in the atmosphere has been
enhanced through deep mixing on the RGB.  However, the results of the
\markcite{snc98}Sweigart \& Catelan (1998) models appear to be rather
sensitive to the assumed mass loss parameters (mean and distribution),
which are at present poorly constrained by either observation or
theory.  Thus, it remains unclear whether the existing observations
can isolate a unique set of physical parameters which explain the
production of the extended blue HB.

We note one additional piece of observational evidence which may be
brought to bear on this problem.  The shapes of the c-type RRL light
curves in NGC 6441 appear to differ significantly from their
``normal'' counterparts.  They have longer phase intervals between
minimum and maximum light, and the minima appear to be sharper.
Perhaps by adjusting the physical parameters in stellar pulsation
models (e.g., \markcite{bc97b}Bono et al. 1997b), similar light curves
can be produced.  If so, the intersection of plausible parameters from
evolution and pulsation theory could further isolate the physical
differences between ``normal'' clusters and those like NGC 6441.  We
encourage additional work on this problem both from the theoretical
and observational perspectives.


\acknowledgements

We thank M\'{a}rcio Catelan, Mario Mateo, Ata Sarajedini, Horace
Smith, and an anonymous referee for helpful comments and suggestions.
Support for this work was provided by NASA through Hubble Fellowship
grant HF-01082.01-96A awarded by the Space Telescope Science
Institute, which is operated by the Association of Universities for
Research in Astronomy, Inc., for NASA under contract NAS 5-26555.
Support was also provided by the National Science and Engineering
Council of Canada through a grant to D.L.W.


\bigskip

\appendix{Appendix:  Notes on Individual Variable Stars}

\noindent {\bf V01--V12:} Variable stars listed in the
\markcite{hsh73}Hogg (1973) catalog (\markcite{cc_97}Clement 1997).
 
\noindent {\bf V11 and V12:} \markcite{hnh76}Hesser \& Hartwick (1976)
identified these stars as possible RR Lyrae variables (RRL).  We
obtained an rms scatter about the mean values in Table 5 of 0.03 mag
in $V$ and $\sim$0.08 mag in $I$ for both candidates.  Their $I_{WS}$
variability indices are 0.0 and 3.2, respectively.  Our data thus show
no sign of variability in these stars.  Indeed, they are $>0.2$ mag
bluer than the {\it bona fide} RRL stars we have discovered (see
Figure 4).

\noindent {\bf SV1:} definitely variable, but the class of variable is
uncertain.  The data fold with minimal scatter at $P=0.3239$ days to
produce the light curve (LC) of a c-type RRL, and at $P=0.6472$ days
to produce the LC of a contact eclipsing binary.  The star is
$\sim$0.8 mag brighter than the member RRL.  

\noindent {\bf SV2:} definitely variable, but the class of variable is
uncertain.  The data fold with minimal scatter at $P=0.5612$ days to
produce an RRL LC, and at $P=1.122$ days to produce the LC of a
contact eclipsing binary.  The RRL LC is too sharp at the extrema for
a typical RRc, and the rise-time is too long for a typical RRab.
However, the star falls among the other cluster RRL in the CMD.

\noindent {\bf SV3:} definitely variable, but the variability class is
uncertain.  The data fold well at $P=0.9016$ days, yielding a contact
eclipsing binary.  A less convincing fold is achieved with $P=0.4508$
days to produce an RRc LC, but the minimum is too sharp and the
maximum is skewed to late phases.  The star's position in Figure 4
suggests it is an RRL member of NGC 6441.

\noindent {\bf SV4:} definitely variable, but the variability class is
uncertain.  The $V$-band data fold well with $P=0.3167$ days, yielding
an RRc LC, though the $I$-band curve has significant scatter.  A
period of $P=0.4814$ days yields a contact eclipsing binary with an
eccentric orbit.  The color is 0.3 mag bluer than the {\it bona fide}
cluster RRL.  

\noindent {\bf SV5:} definitely variable, but the variability class is
uncertain.  The data fold well with $P=0.3615$ days, yielding an RRc
LC, though the $I$-band curve has significant scatter.  A period of
$P=0.7230$ days yields a contact eclipsing binary with considerable
asymmetry.  The star's position in Figure 4 suggests it is an RRL
member of NGC 6441.  




 
\begin{deluxetable}{rrrcccc}
\tablewidth{29pc}
\tablecaption{Photometry of 16,011 Stars toward NGC 6441\tablenotemark{a}}
\tablehead{
\colhead{ID}                 & \colhead{$X_{\rm pix}$}    &
\colhead{$Y_{\rm pix}$}      & \colhead{$V$}              & 
\colhead{$\sigma_{\rm V}$}   & \colhead{$V-I$}            & 
\colhead{$\sigma_{\rm V-I}$}
}
\startdata  
 1 & 1267.64 &  -27.67 & 17.001 &  0.169 &  1.071 &  0.192 \nl
 2 & 1336.58 &  -25.93 & 18.611 &  0.043 &  1.150 &  0.059 \nl
 3 &  949.77 &  -25.65 & 17.553 &  0.151 &  1.567 &  0.171 \nl
 4 & 1225.50 &  -25.00 & 18.059 &  0.021 &  1.206 &  0.030 \nl
 5 & 1216.89 &  -24.55 & 18.799 &  0.041 &  1.135 &  0.069 \nl
\enddata
\tablenotetext{a}{Table 1 is available in its entirety, in 
electronic format, at
http://bethe.bgsu.edu/$\sim$layden/ASTRO/PREPRINTS/preprints.html}
\end{deluxetable}

 
\begin{deluxetable}{lrrcccccc}
\tablewidth{34pc}
\tablecaption{Mean Photometry of Long Period Variable Stars}
\tablehead{
\colhead{Star}               & \colhead{$X_{\rm pix}$}    &
\colhead{$Y_{\rm pix}$}      & \colhead{${\overline{I}}_{\rm May}$}    &
\colhead{${\overline{V-I}}_{\rm May}$}    & \colhead{$N_{\rm May}$}    &
\colhead{${\overline{I}}_{\rm Jun}$}      & \colhead{${\overline{V-I}}_{\rm Jun}
$}  &
\colhead{$N_{\rm Jun}$}
}
\startdata
V13 & 1005 & 1185 & 12.98 & 3.49 &  5 & 12.59 & 3.05 &  18 \nl  
V1\tablenotemark{a}&  890 & 1112 & 13.76 & 4.97 &  6 & 12.85 & 4.20 &  17 \nl
V14 &  432 & 1183 & 12.89 & 2.70 &  5 & 13.27 & 3.17 &  20 \nl  
V15 & 1156 & 1609 & 13.03 & 3.85 &  5 & 12.79 & 3.54 &  18 \nl  
V16 & 1360 & 1147 & 13.16 & 3.65 &  5 & 12.90 & 3.12 &  18 \nl  
V17 & 1182 &  937 & 13.19 & 2.88 &  5 & 13.43 & 3.05 &  19 \nl  
V18 & 1463 &   49 & 12.78 & 3.90 &  5 & 13.04 & 4.09 &  19 \nl  
V19 & 1065 & 1729 & 13.32 & 6.60 &  4 & 12.83 & 6.28 &  13 \nl  
V20 &  477 & 1510 & 13.03 & 2.70 &  5 & 13.29 & 2.95 &  21 \nl  
V21 & 1423 & 1401 & 13.02 & 4.53 &  5 & 12.81 & 4.39 &  17 \nl  
V22 & 1268 & 1076 & 12.95 & 2.95 &  5 & 12.86 & 2.82 &  16 \nl  
V23 &  979 & 1129 & 12.80 & 3.46 &  5 & 12.68 & 3.29 &  16 \nl  
V24 &  480 &  962 & 13.07 & 3.06 &  5 & 13.17 & 3.19 &  17 \nl  
V25 & 1431 & 1347 & 14.68 & 4.10 &  7 & 14.44 & 4.18 &  19 \nl  
V26 & 1271 &  747 & 12.76 & 3.75 &  5 & 12.67 & 3.54 &  17 \nl  
V27 &  823 &  396 & 12.98 & 3.02 &  5 & 12.93 & 2.97 &  17 \nl  
V28 & 1092 &  190 & 14.33 & 2.44 &  7 & 14.29 & 2.37 &  21 \nl  
V9\tablenotemark{a}& 1073 & 1119 & 12.52 & 2.84 &  3 & 12.23 & 2.53 &  12 \nl
V29 & 1486 &  501 & 12.46 & 4.80 &  3 & 12.81 & 4.97 &  17 \nl  
V30 &  354 &  658 & 12.38 & 3.08 &  3 & 12.26 & 2.93 &  12 \nl  
V31 &  804 & 1581 & 12.14 & 3.02 &  3 & 12.25 & 3.12 &  12 \nl  
V32 & 1062 & 1045 & 12.94 & 2.76 &  3 & 12.87 & 2.64 &  12 \nl  
V33 &  936 &  986 & 13.00 & 2.56 &  5 & 13.15 & 2.70 &  13 \nl  
V34 &  601 & 1580 & 13.31 & 3.26 &  4 & 13.23 & 3.14 &  19 \nl  
V35 &  877 & 1017 & 13.52 & 2.19 &  5 & 13.39 & 2.11 &  19 \nl  
V6\tablenotemark{a}&  933 &  877 & 13.56 & 1.66 &  5 & 12.79 & 1.40 &  18 \nl
V2\tablenotemark{a}&  918 &  938 & 12.09 & 2.70 &  3 & 13.32 & 4.25 &   4 \nl
V3\tablenotemark{a}&  124 & 1212 & 11.60 & 3.49 &  3 & 12.37 & 4.74 &  12 \nl
V4\tablenotemark{a}&  853 & 1440 & 13.00 & 2.88 &  5 & 13.00 & 2.90 &  18 \nl
V5\tablenotemark{a}&  504 &  425 & 12.29 & 3.93 &  3 & 11.54 & 2.98 &   4 \nl
V10\tablenotemark{a}&  819 & 1149 & 12.72 & 3.03 &  4 & 12.92 & 3.39 &  18
\enddata
\tablenotetext{a}{see comment in Appendix.}
\end{deluxetable}

 
\begin{deluxetable}{lrrllllll}
\tablewidth{40pc}
\tablecaption{Photometry of RR Lyrae Variables}
\tablehead{
\colhead{Star}               & \colhead{$X_{\rm pix}$}    &
\colhead{$Y_{\rm pix}$}      & \colhead{Period}           &
\colhead{$<V>$}              & \colhead{$<I>$}            & 
\colhead{$\Delta V$}         & \colhead{$(V-I)_{\rm min}$}  &
\colhead{Comment}
}
\startdata
V36 & 1703 & 1299 & 0.5078 & 16.54 & 15.50 & 1.16 & 1.16 & bulge  \nl   
V37 &  926 &  705 & 0.6132 & 17.38 & 16.41 & 1.04 & 1.10 &        \nl   
V38 &  994 &  605 & 0.7347 & 17.35 & 16.33 & 0.76 & 1.10 &        \nl   
V39 &  716 &  847 & 0.8330 & 17.53 & 16.36 & 0.72 & 1.28 &        \nl   
V40 & 1088 & 1139 & 0.6490 & 17.22 & 16.21 & 1.11 & 1.18 &        \nl   
V41 &  923 & 1131 & 0.7345 & 16.65 & 15.26 & 0.38 & 1.46 & see Appendix\nl 
V42 & 1011 &  765 & 0.8140 & 17.39 & 16.33 & 0.57 & 1.14 &        \nl   
V43 &  897 &  861 & 0.7730 & 17.38 & 16.34 & 0.52 & 1.10 &        \nl   
V44 &  975 & 1138 & 0.6090 & 16.58 & 15.23 & 0.64 & 1.45 & see Appendix\nl 
V45 & 1179 & 1339 & 0.5028 & 17.15 & 16.11 & 0.73 & 1.09 &        \nl   
V46 & 1377 & 1198 & 0.9050 & 17.31 & 16.18 & 0.42 & 1.18 &              
\enddata
\end{deluxetable}

 
\begin{deluxetable}{lrrllllll}
\tablewidth{35pc}
\tablecaption{Photometry of Eclipsing Variables}
\tablehead{
\colhead{Star}               & \colhead{$X_{\rm pix}$}    &
\colhead{$Y_{\rm pix}$}      & \colhead{Period}           &
\colhead{$V_{\rm max}$}      & \colhead{$I_{\rm max}$}    & 
\colhead{$\Delta V$}         & \colhead{$\Delta I$}       &
\colhead{Comment}
}
\startdata
V47 &  836 & 1765 & 0.703  & 16.20 & 15.02 & 1.47 & 1.10 & detatched \nl 
V48 &  158 & 1171 & 0.6674 & 15.23 & 14.22 & 0.32 & 0.28 & contact   \nl 
V49 &  365 &  768 & 1.010  & 16.52 & 15.45 & 0.36 & 0.20 & detatched?\nl 
V50 &  838 & 1410 & 0.4335 & 17.85 & 16.55 & 0.55 & 0.55 & contact?   
\enddata
\end{deluxetable}

\begin{deluxetable}{crrccl}
\tablewidth{32pc}
\tablecaption{Photometry of Suspected Variable Stars}
\tablehead{
\colhead{Star}               & \colhead{$X_{\rm pix}$}    &
\colhead{$Y_{\rm pix}$}      & \colhead{${\overline{V}}$}            & 
\colhead{${\overline{V-I}}$}            & \colhead{Comment}
}
\startdata
SV1 & 1364 & 1501 & 15.81 &  0.99 & see Appendix            \nl 
SV2 &  536 & 1057 & 17.22 &  1.03 & see Appendix            \nl 
SV3 &  580 & 1244 & 17.25 &  1.07 & see Appendix            \nl 
SV4 & 1268 & 1129 & 17.24 &  0.72 & see Appendix            \nl 
SV5 &  912 &  796 & 17.37 &  0.97 & see Appendix            \nl 
SV6 & 1062 &  940 & 16.89 &  2.59 & crowded, LPV?           \nl 
SV7 & 1034 &  969 & 15.13 &  2.49 & crowded, LPV?           \nl 
SV8 &  643 &  563 & 17.21 &  1.18 & blended image, RR Lyrae?  \nl 
SV9 & 1359 &   93 & 17.10 &  1.17 & blended image, eclipsing binary? \nl 
V11 & 1120 &  755 & 17.86 &  0.48 & non-variable, see Appendix \nl 
V12 & 1123 &  867 & 17.29 &  0.69 & non-variable, see Appendix \nl 
\enddata
\end{deluxetable}

 
\begin{deluxetable}{rcccccr}
\tablewidth{30pc}
\tablecaption{Time Series Photometry of Variable Stars\tablenotemark{a}}
\tablehead{
\colhead{Star}                 & \colhead{$HJD$}    &
\colhead{$V$}                  & \colhead{$\sigma_{\rm V}$}   & 
\colhead{$I$}                  & \colhead{$\sigma_{\rm V-I}$} &
\colhead{$Q$\tablenotemark{b}}
}
\startdata
V1 & 258.6055 & 17.231 &  0.010 & 12.962 &  0.005 & 3 \nl
V1 & 258.6006 & 17.195 &  0.021 & 12.901 &  0.004 & 4 \nl
V1 & 258.8444 & 17.201 &  0.016 & 12.920 &  0.005 & 2 \nl
V1 & 258.8511 & 17.164 &  0.010 & 12.909 &  0.004 & 1 \nl
V1 & 259.5721 & 17.142 &  0.013 & 12.906 &  0.005 & 4 \nl
... &          &        &        &        &        &   \nl
V1 & 265.8469 & 17.371 &  0.063 & 12.847 &  0.007 & 3 \nl
V1 & 266.7073 & 16.844 &  0.021 & 12.785 &  0.010 & 4 \nl
V2 & 258.8444 & 17.352 &  0.033 & 13.217 &  0.009 & 2 \nl
V2 & 225.6801 & 14.754 &  0.017 & 12.078 &  0.012 & 2 \nl
V2 & 260.7804 & 17.512 &  0.034 & 13.300 &  0.012 & 2
\enddata
\tablenotetext{a}{Table 6 is available in its entirety, in electronic format,
at http://bethe.bgsu.edu/$\sim$layden/ASTRO/PREPRINTS/preprints.html}
\tablenotetext{b}{Quality Code, $Q$.  See \S4.3 for definition.}
\end{deluxetable}


\begin{deluxetable}{cccccccc}
\tablewidth{35pc}
\tablecaption{Reddening Toward NGC 6441}
\tablehead{
\colhead{Sector}     & 
\colhead{$\alpha_1$\tablenotemark{a}}    &
\colhead{$\alpha_2$\tablenotemark{a}}    & 
\colhead{$Med(V-I)^{13.6}_{13.9}$}       &
\colhead{$N_{\rm sec}$}      & \colhead{RR Lyr}          & 
\colhead{$E(V-I)$}           & \colhead{$E(B-V)$}          
}
\startdata
1 &  0    & 45    & 1.391 &  64 &V36\tablenotemark{b}& 0.588 & 0.456 \nl
  &       &       &       &     &V46\tablenotemark{b}& 0.612 & 0.474 \nl
2 & 45    & 90    & 1.429 &  63 & V40 & 0.604 & 0.468 \nl
  &       &       &       &     & V45 & 0.516 & 0.400 \nl
3 & 90    & 135   & 1.472 &  59 & V41 & 0.884 & 0.683 \nl
  &       &       &       &     & V44 & 0.874 & 0.676 \nl
4 & 135   & 180   & 1.524 &  95 & ~-  &       &       \nl
5 & 180   & 225   & 1.460 &  82 & V39 & 0.707 & 0.548 \nl
6 & 225   & 270   & 1.401 &  79 & V37 & 0.529 & 0.410 \nl
  &       &       &       &     & V38 & 0.531 & 0.412 \nl
  &       &       &       &     & V42 & 0.572 & 0.444 \nl
  &       &       &       &     & V43 & 0.532 & 0.413 \nl
7 & 270   & 315   & 1.409 &  92 & ~-  &       &       \nl
8 & 315   & 360   & 1.393 &  90 & ~-  &       &       
\enddata
\tablenotetext{a}{These angles (in degrees) define the sector
boundaries, see \S5.2.}
\tablenotetext{b}{V36 and V46 lie outside Sector 1, at radii of 300
and 165 arsec, respectively.}
\end{deluxetable}


\begin{deluxetable}{ccc}
\tablewidth{30pc}
\tablecaption{Star Counts in NGC 6441}
\tablehead{
\colhead{Region\tablenotemark{a}} & \colhead{Cluster}    &
\colhead{Field}
}
\startdata
BHB                 & 175 & ~83 \nl
RRL                 & ~~~9& ~~~0\nl
RC                  & 642 & ~41 \nl
RGB                 & 918 & 238 \nl
LPV                 & ~10 & ~~~0\nl
RHB\tablenotemark{b}& 107 & ~60
\enddata
\tablenotetext{a}{CMD regions for BHB, RC, and RGB counts are defined in
Fig. 11.}
\tablenotetext{b}{Red HB region is the BHB region redward of $(V-I)_0
= 0.49$ mag.}
\end{deluxetable}


\clearpage

\bigskip
 
\begin{center}
{\bf Figure Captions}
\end{center}
 
\medskip


\noindent {\bf Figure 1:}  A 30 sec exposure image of NGC 6441 in
the $V$-band.  North is down and East is to the left.  The field is
$13.5\arcmin$ across.  Pixel coordinates range from ($X$, $Y$) = (1,
1) at the lower-left to (2048, 2048) at the upper right, with the
cluster center lying at (1014, 1000).  The bright star HR 6630 has
been masked out of the image.  A larger scale image is available at
http://bethe.bgsu.edu/~layden/ASTRO/PREPRINTS/preprints.html. 

\medskip
 

\noindent {\bf Figure 2:}  Color-magnitude diagrams of stars toward
(a) NGC 6441, and (b) a field centered $13.3\arcmin$ North of the
cluster.  In both panels, only stars with small errors ($\sigma_V <
0.050$ mag and $\sigma_{V-I} < 0.071$ mag) are shown.  In (a), only
stars lying within $440\arcsec$ of the cluster center are shown, while
in (b) only stars within $400\arcsec$ are shown.  The line segment
represents the reddening vector ($A_V = 1.0$ mag).

\medskip
 

\noindent {\bf Figure 3:}  Color-magnitude diagrams for stars with
moderate errors ($\sigma_V < 0.10$ mag and $\sigma_{V-I} < 0.14$ mag)
and located (a) within $440\arcsec$ of NGC 6441, (b) within
$400\arcsec$ of the center of the adjacent field, (c) between 40 and
$160\arcsec$ of NGC 6441.  In (d), the points in (c) are shown after
the statistical subtraction of the CMD generated from off-cluster
stars located in the same radial zone. 

\medskip
 

\noindent {\bf Figure 4:}  The NGC 6441 color-magnitude diagram from
Fig 3a with the variable stars marked as follows: long-period
variables from Table 2 ({\em solid squares}); RR Lyrae stars from Table 3
($\bullet$); eclipsing binaries from Table 4 ($\bigtriangleup$);
variables of uncertain class (SV1--SV5) from Table 5 ($\ast$); and the
remaining suspected variables from Table 5 ($\times$). 

\medskip
 

\noindent {\bf Figure 5:}  Mean light curves for the blue variable
candidates which yielded clear periods.  Each panel shows the $V$
(lower curve) and $I$ (upper curve, in some panels offset vertically
by the indicated amount for convenience of display) light curve for
the indicated star and period.  In all panels, the minor tick marks
indicate intervals of 0.2 mag.  

\medskip


\noindent {\bf Figure 6:}  Mean light curves for the blue variable
candidates which yielded two possible periods.  In each panel, the
left-most set of light curves ($V$ below; $I$ above, possibly shifted
vertically) employs the shorter period and has the characteristics of
a c-type RR Lyrae, while the right-most set employs the longer period,
and resembles a contact binary. 

\medskip
 

\noindent {\bf Figure 7:}  Finder charts for the variable stars.  Each
image is $20\arcsec$ square, and is oriented with North at the top and
East to the left.  For the LPVs (V1--V35 plus SV6 and SV7), $I$-band
images are used, while $V$-band images are used for all other stars.

\medskip
 

\noindent {\bf Figure 8:}  The cumulative radial distributions of
stars of various classes, normalized to the total number inside
$440\arcsec$ radius for: red clump stars ({\em solid line}, RC); blue
HB stars ({\em dotted line}, BHB); RR Lyrae stars ({\em short-dashed
line}, RRL); LPV stars ({\em long-dashed line}); and eclipsing
binaries ({\em dash-dot line}, ECL).  The $V$-band light distribution
(sky subtracted, in intensity units) indicates the distribution
expected for cluster members ($\Box$, CL).  A uniform
illumination over the useable CCD pixels indicates the distribution
expected for a field star population ($\times$, FLD).  The radial
locations of the suspected variable stars, SV1--SV9, are marked along
the top, with the probable RR Lyrae stars in italics.

\medskip
 

\noindent {\bf Figure 9:}  The Period-Amplitude diagram for
metal-rich RR Lyrae stars.  The small symbols include ab-type field
RRL with $[Fe/H] > -0.5$ ($\times$), field RRL with $-1.0 < [Fe/H] <
-0.5$ ($\ast$), and ab- and c-type RRL in NGC 6171
(\markcite{dic70}Dickens 1970, $\Box$).  Large symbols include: NGC
6441 variables V36, V41, and V44 ($\ast$); SV2, SV3, and SV5
($\circ$); and the eight other RRL from Table 3 ($\bullet$); 47 Tuc
variable V9 (RRab, {\em solid square}); NGC 6388 variables V28 (RRab,
{\em solid triangle}) and V27 (RRc, $\bigtriangleup$) from
\markcite{ssb94}Silbermann et al. (1994).  The dotted region
represents one of the model predictions of \markcite{snc98}Sweigart \&
Catelan (1998, from their Figure 3).

\medskip
 

\noindent {\bf Figure 10:}  Generalized color histograms of red clump
stars in the eight sectors numbered in the figure (see Table 7).  Each
histogram is a sum of unit-area Gaussian curves placed at the $V-I$
color of each star and having a width equal to the color error of the
star.  The shifts between sectors indicate the presence and degree of
differential reddening.

\medskip
 

\noindent {\bf Figure 11:}  The differentially de-reddened
color-magnitude diagram of NGC 6441.  Variable stars are indicated as
follows: long period variables ({\em filled square}), ab-type RRL
($\bullet$), suspected c-type RRL ($\ast$).  The dotted boxes define
the regions used to count blue HB stars (BHB), red clump stars (RC),
and red giant branch stars (RGB) in Sec. 5.5.

\end{document}